\documentclass[10pt,twocolumn,letterpaper]{article}

\usepackage{iccv}
\usepackage{times}
\usepackage{epsfig}
\usepackage{graphicx}
\usepackage{amsmath}
\usepackage{amssymb}
\usepackage [english]{babel}
\usepackage [autostyle, english = american]{csquotes}
\MakeOuterQuote{"}

\usepackage[pagebackref=true,breaklinks=true,letterpaper=true,colorlinks,bookmarks=false]{hyperref}

\hypersetup{
 colorlinks=true,
 citecolor=blue
}

\iccvfinalcopy 

\ificcvfinal\fi

\begin{document}

\title{Noise2Fast: Fast Self-Supervised Single Image Blind Denoising}

\author{Jason Lequyer$^{1,2}$\\
{\small jlequyer@lunenfeld.ca}
\and
Reuben Philip$^{1,2}$\\
{\small rphilip@lunenfeld.ca}
\and
Amit Sharma$^{1}$\\
{\small asharma@lunenfeld.ca}
\and
Laurence Pelletier$^{1,2}$*\\
{\small pelletier@lunenfeld.ca}
\and
1. Lunenfeld-Tanenbaum Research Institute\\
{\small 600 University Avenue, Toronto, Ontario, Canada}\\
{\small *Corresponding author \qquad \qquad\qquad\qquad\qquad\qquad\qquad\qquad}\\
\and
2. Department of Molecular Genetics\\
{\small University of Toronto, Toronto, Ontario, Canada}\\
}

\maketitle
\ificcvfinal\thispagestyle{empty}\fi

\begin{abstract}
In the last several years deep learning based approaches have come to dominate many areas of computer vision, and image denoising is no exception. Neural networks can learn by example to map noisy images to clean images. However, access to noisy/clean or even noisy/noisy image pairs isn't always readily available in the desired domain. Recent approaches have allowed for the denoising of single noisy images without access to any training data aside from that very image. But since they require both training and inference to be carried out on each individual input image, these methods require significant computation time. As such, they are difficult to integrate into automated microscopy pipelines where denoising large datasets is essential but needs to be carried out in a timely manner. Here we present Noise2Fast, a fast single image blind denoiser. Our method is tailored for speed by training on a four-image  dataset produced using a unique form of downsampling we refer to as "checkerboard downsampling". Noise2Fast is faster than all tested approaches and is more accurate than all except Self2Self, which takes well over 100 times longer to denoise an image. This allows for a combination of speed and flexibility that was not previously attainable using any other method.

\end{abstract}

\section*{Main}

Image noise is the random fluctuation of color or intensity values that is inherent to image acquisition. It usually presents as a hazy shroud that obscures an otherwise clear visual signal. Image denoising methods try to fix this by removing noise after the fact, usually by exploiting the innate structure and pattern of the underlying signal and leveraging it against the apparent stochasticity of the noise. Denoising is particularly important in live cell imaging applications, where a balance between the conflicting considerations of resolution, phototoxicity and throughput can force experimenters to accept a considerable amount of noise as necessary to achieving their goals. 

Many techniques are focused on explicitly modelling noise, based on an understanding of its origin; for example it is known that confocal microscopy is mainly subject to a combination of Gaussian and Poisson distributed noise \cite{Confoc5}. However, with the advent of deep learning, such explicit models are avoidable by instead training a neural network to learn how to map noisy images to their clean counterparts, such as in DnCNN \cite{Zhang_2017}, or even just by training it to map noisy pairs of images to one another, such as in Noise2Noise \cite{pmlr-v80-lehtinen18a}. But, trained methods like these cannot be expected to perform well on image types that were not well represented in the training set. In cases where we do not have access to such training data, alternatives must be considered.

One such alternative is Noise2Void \cite{Krull_2019_CVPR}. Noise2Void denoises images by using a masking procedure wherein the neural network learns to fill in pixel gaps in the noisy image. The network's failure to learn the noise causes it to denoise the underlying image. Although it was trained on entire datasets of images with similar noise levels in their paper, Noise2Void can be adapted to denoise single noisy images based purely on the information contained within that image, appealing to no outside information or pretrained weights. This basic process was improved and generalized in Noise2Self \cite{pmlr-v97-batson19a} and further refined in Self2Self \cite{Quan_2020_CVPR} to achieve single image denoising results that are competitive with traditional fully trained methods. Recently, an approach based on corrupting the input image into pairs of new noisy realizations called Recorrupted2Recorrupted (R2R) \cite{recor} has emerged and achieved better results than Self2Self on real world noise. However, all viable single image denoisers to date require a considerable amount of time to run, making them impractical for use on high resolution microscopy images in time sensitive situations. 

To alleviate this, we propose Noise2Fast. Noise2Fast is similar to the masking based methods in that the network is blind to many of the input pixels during training. Our method is inspired by a recently published approach called Neighbor2Neighbor \cite{Ne2Ne} where the neural network learns a mapping between adjacent pixels. We tune our method to speed by using a discrete four image training set obtained by an unusual form of downsampling we refer to as "checkerboard downsampling" and train a fairly small neural network on this discrete training set, validating using the original full-sized image to determine convergence. Noise2Fast is faster than all compared methods, and is more accurate than all tested methods except for Self2Self, which takes well over 100 times as long to denoise a single image.

\subsubsection*{Theoretical Background}
\begin{figure*}
\begin{center}
\includegraphics[width=1\linewidth]{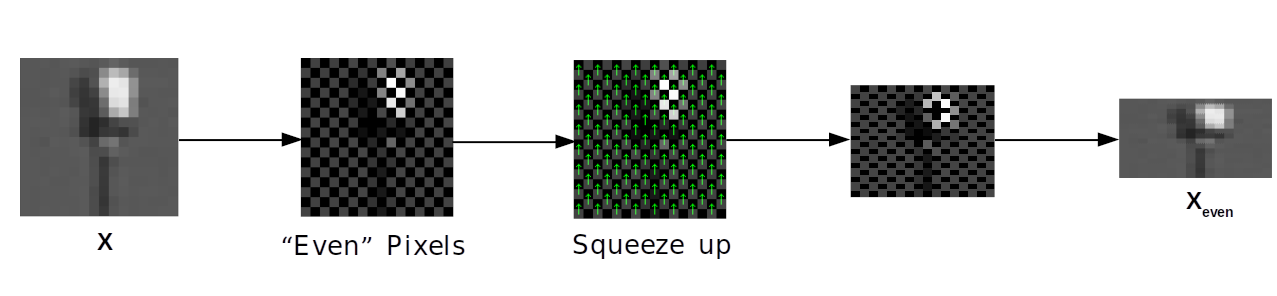}
\end{center}
\caption{Checkerboard downsampling illustrated. We take our initial image, remove one half of all pixels in a checkerboard pattern, and shift the remaining pixels to fill in the gaps left behind.}
\end{figure*}
Consider a 2D image $\bf{x} \in \mathbb{R}^{mxn}$ composed of both signal and noise $\bf{s},\bf{n}\in \mathbb{R}^{mxn}$. That is to say
\begin{align} \mathbf{x} = \mathbf{s} + \mathbf{n}. \end{align}
Denoising is concerned with the inverse problem of inferring $\mathbf{s}$ from $\bf{x}$ (or equivalently inferring $\mathbf{n}$ and then solving for $\bf{s}$). A neural network attempts to solve this problem by finding a function $f_\theta:\mathbb{R}^{mxn} \rightarrow \mathbb{R}^{mxn}$ (parameterized by the network weights $\theta$) such that
\begin{align} f_\theta(\mathbf{x}) \approx \mathbf{s}. \end{align}

The most intuitive way to train such a network is by using pairs of noisy/clean images and having the network learn a mapping from one to the other. Noise2Noise takes an alternate approach by training the network to learn a mapping from different noisy shots of the same image, allowing for training in the absence of clean ground truth data. Specifically, given two noisy realizations of the same underlying signal $\mathbf{s} + \mathbf{n}_1$ and $\mathbf{s} + \mathbf{n}_2$ Noise2Noise attempts to learn the mapping
\begin{align} \label{eqn:N2N} f_\theta(\mathbf{\mathbf{s}+\mathbf{n}_1}) \rightarrow \mathbf{s}+\mathbf{n}_2. \end{align}
However if we assume mean-zero noise and choose a sensible loss function \cite{pmlr-v80-lehtinen18a}, the network may fail to actually learn the noise $\mathbf{n}_2$, and we will be left with
\begin{align} \label{eqn:N2N2} f_\theta(\mathbf{\mathbf{s}+\mathbf{n}_1}) \approx \mathbf{s}, \end{align}
denoising the image as a result. Although elegant, this method still requires pairs of noisy images to train on. 

Recently, interest has grown in methods that can denoise single noisy images, without this added requirement. To fully understand these methods, we need to adopt a different perspective of how neural networks denoise images. 

Here, we take the view of Krull et al. \cite{Krull_2019_CVPR}, based on the concept of receptive fields. The receptive field of a fully convolutional neural network (FCN) is the set of input pixels that were taken into consideration for a given output pixel prediction. For example, in our above scenario suppose $(i,j) \in \mathbb{N}_{\leq m} \times \mathbb{N}_{\leq n}$ are the co-ordinates of some pixel in the output image  $f_\theta(\bf{x})$. Then the receptive field of that pixel is the set of indices $RF(i,j) \subseteq \mathbb{N}_{\leq m} \times \mathbb{N}_{\leq n}$ such that $f_\theta(\bf{x})_{(i,j)}$ depends only upon the value of $\left.\bf{x}\right|_{RF(i,j)}$ (typically this will be a small square patch of the image $\bf{x}$). We can then view the neural network as a mapping from the input image along some receptive field to its corresponding output pixel, with the goal of finding $\theta$ such that
\begin{align} f_\theta \big(\left.\mathbf{x}\right|_{RF(i,j)} \big) \approx \mathbf{s} \big(i,j\big), \end{align}
for every $(i,j) \in \mathbb{N}_{\leq m} \times \mathbb{N}_{\leq n}$. The question though, is how to train these networks without any actual training data other than the noisy image itself. Blind-spot methods approach this by excluding the center pixel from the receptive field (either by removing/replacing it \cite{pmlr-v97-batson19a, Krull_2019_CVPR} or ignoring it altogether using partial convolutions \cite{Quan_2020_CVPR}), and training the network to recover this center pixel from its surroundings. More specifically, they train the network to learn the mapping
\begin{align} \label{eqn:RF} f_\theta \big(\left.\mathbf{x}\right|_{RF(i,j) \setminus (i,j)}  \big) \rightarrow \ \mathbf{x} \big(i,j\big). \end{align}
However, just as in Noise2Noise, the network fails to learn the noise, leaving us with
\begin{align} \label{eqn:RFR} f_\theta \big(\left.\mathbf{x}\right|_{RF(i,j) \setminus (i,j)}  \big) \approx \ \mathbf{s} \big(i,j\big). \end{align}

Excluding the center pixel is crucial and ensures that the network does not just learn the identity. However a side effect of this is that the neural network does not give proper weight to the pixel itself when computing the output, which is unfortunate, since the pixel itself is always going to be the best individual predictor of its denoised value.

Our method takes a related, but slightly different approach. Instead of masking the input image, we explicitly divide the input image in two, by using a simple downsampling method that we refer to here as "checkerboard downsampling". This process is easier to visualize than explain (see Figure 1), however we take our input image $\mathbf{x}$ and split it into two smaller images composed of the even pixels (where $i+j$ is even) and odd pixels (where $i+j$ is odd) respectively, and compress them into the two following $m \times \frac{1}{2}n$ images
\begin{align} \mathbf{x}_\text{even}(i,j) &= \mathbf{x}(i,2j+(i \;\mathrm{mod}\; 2)), \\
\mathbf{x}_\text{odd}(i,j) &= \mathbf{x}(i,2j+(i \;\mathrm{mod}\; 2)+1). \end{align}
We can call these the "up" checkerboard downsamples, since they involve shifting everything up one pixel to close the image. This type of downsampling is highly susceptible to aliasing, however the goal here isn't visual clarity, it's to preserve the original noise model of our image as much as possible. 

Now suppose we train our neural network to learn the mapping
\begin{align}\label{eqn:N2F0} f_\theta(\mathbf{x}_\text{even}) \rightarrow \mathbf{x}_\text{odd}. \end{align}
We can rewrite this as
\begin{align} \label{eqn:N2F} f_\theta(\mathbf{s}_\text{even} + \mathbf{n}_\text{even}) \rightarrow \mathbf{s}_\text{even} + \mathbf{n}_\text{odd} + (\mathbf{s}_\text{odd} - \mathbf{s}_\text{even}). \end{align}
Notice that this is analogous to Noise2Noise \ref{eqn:N2N}, except for the addition of the $(\mathbf{s}_\text{odd} - \mathbf{s}_\text{even})$ term. However for every $(i,j) \in \mathbb{N}_{\leq m} \times \mathbb{N}_{\leq n}$, we have that $\mathbf{s}_\text{odd}(i,j)$ and $\mathbf{s}_\text{even}(i,j)$ are adjacent pixels in the original image signal, it is therefore reasonable to think this term would be very small in all but the most highly dynamic regions. Indeed, in our testing, we found that even if we cheat and subtract out the term using known ground truth values, there was no measurable gain in denoising performance. We therefore claim that for most natural images,
\begin{align} \label{eqn:ODD} \mathbf{s}_\text{even} + \mathbf{n}_\text{odd} + (\mathbf{s}_\text{odd} - \mathbf{s}_\text{even}) \approx \mathbf{s}_\text{even} + \mathbf{n}_\text{odd}. \end{align}
Then, analogous with Noise2Noise ($\ref{eqn:N2N}, \ref{eqn:N2N2}$), training our network as outlined in $\ref{eqn:N2F0}$ should, in effect, find weights $\theta$ such that
\begin{align} f_\theta(\mathbf{x}_\text{even}) \approx \mathbf{s}_\text{even}. \end{align}
However, in our experiments we have witnessed a much stronger result than this. In particular, we observe that a network trained as in $\ref{eqn:N2F0}$ will not just learn to denoise the downsampled image, but the entire image as a whole. That is
\begin{align} f_\theta(\mathbf{x}) \approx \mathbf{s}. \end{align}
To explain this phenomenon, we return to the receptive field based perspective of $\ref{eqn:RF}$. In this case, our network is trained to learn the mapping

\begin{align} \label{eqn:RF5} f_\theta \big(\left.\mathbf{x}_\text{even}\right|_{RF(i,j)}  \big) \rightarrow \ \mathbf{x}_\text{odd} \big(i,j\big). \end{align}

It is known, and is often exploited by denoising algorithms, that single images contain significant internal redundancy in the form of recurrent patches \cite{5995401}. It is also known, and is crucial to some super-resolution methods, that single images have a certain degree of self-similarity, and hence these patches also recur across scales \cite{5459271,7299621}. This across-scale patch recurrence implies a level of commonality between the two sets 
\begin{align} \{ \left.\mathbf{x}_\text{even}\right|_{RF(i,j)} : (i,j) \in \mathbb{N}_{\leq m} \times \mathbb{N}_{\leq n}\}, \\
\{ \left.\mathbf{x}\right|_{RF(i,j)} : (i,j) \in \mathbb{N}_{\leq m} \times \mathbb{N}_{\leq n}\}. \end{align}
Hence, a neural network trained to learn $\ref{eqn:N2F0}$ may be applicable to the overarching denoising task. 

Our method uses this basic principle to generate a small training set of four image pairs (each one produced by a form of checkerboard downsampling). This compact training set allows for rapid network convergence and hence quick single image denoising results that were previously unattainable with such a high degree of accuracy. 

\subsubsection*{Contribution and Significance}

We present a training scheme and accompanying neural network for blind single-shot denoising. Our method uses only information contained within the single noisy input image to train its weights. Our two main contributions are as follows:
\begin{itemize}
\item{{\bf A novel denoising method which combines an unusual downsampling method with Neighbor2Neighbor.}} Our method uses checkerboard downsampling to generate a small fixed dataset for rapid convergence. We then apply our network trained on this smaller dataset to denoise the larger input image. 
We also use the larger input image to validate our network and determine when it has reached maximum accuracy, which is necessary because Noise2Noise based methods can start to overfit at a certain point and performance drops. 
To our knowledge, this procedure as a whole is novel; in particular, we
are unaware of another denoising method that makes use of this unusual form of downsampling, nor are we aware of any other single image blind denoisers that attempt to build a small fixed training set of images. We are also unaware of any other method that successfully employs a one-size-fits all validation strategy to avoid overfitting, among methods that are susceptible to this.
\item{{\bf High accuracy and significant speed gains over existing methods.}} Our method is tailored specifically for speed; using a small four image dataset ensures rapid convergence, while the theoretical underpinnings of our method ensure its accuracy. In terms of PSNR and SSIM, the only tested method more accurate than the one we propose here is Self2Self which is 300-700 times slower, for example it requires an average of over 3 hours to denoise a 512x512 image that our method can denoise in under 45 seconds on an RTX 5000 mobile GPU. This large time investment makes it impractical for on-demand usage in larger image screens wherein denoising is but one step in a much bigger process. \end{itemize}

Single-shot blind denoisers are a valuable tool for their convenience and their broad applicability. However, thus far, the only such tool that achieves accuracy comparable to trained methods is Self2Self and possibly R2R \cite{recor}, which both require a massive amount of time to run on modern technology. Here, we present a much faster alternative, with only a small drop in accuracy.

In particular, the speed of our tool makes it usable in smart microscopy pipelines (e.g. \cite{Hasle2020, Kanfer2021, Yan2021}) where the microscope captures an image and then responds to information contained within that captured image (e.g. zoom in on any cells undergoing mitosis in a given field of view), for which denoising is typically the first step to ensure better classification by the downstream neural networks and/or manual analysis. In such pipelines, time is of the essence since there is only a small window to do analysis, and gathering training data for traditional training based denoising methods can be expensive both in terms of cost and the delays it imposes on research. Additionally, our tool allows for the processing of very large datasets, such as 3D time lapse videos of cells, with a compute overhead 100s of time smaller. 

\subsubsection*{Related Work}
\textbf{Methods that require a training set:} The first attempt to apply Convolutional Neural Networks (CNNs) to the task of denoising was in \cite{firstDn}. This was heavily refined in both the works of Mao et al. \cite{OtherDn} and Zhang et al. \cite{Zhang_2017} (DnCNN) to achieve performance that is still competitive today. Zhang et al. later released FFDNet \cite{FFD}, a denoising CNN designed with speed in mind which, similar to our method, also uses downsampling, although in a different manner and to an entirely different end (see \cite{Sub-pixel}).

The main benefit of using trained methods, outside of their outstanding performance, is that they don't require either implicit or explicit modeling of the type and structure of the noise, they can simply be trained on noisy/clean pairs of images from the desired domain. However, their reliance on noisy/clean image pairs, either real or synthetically generated, can be considered a limitation in situations where we do not have access to ground truth images to train on.

To overcome this limitation, Noise2Noise was developed \cite{pmlr-v80-lehtinen18a}. Noise2Noise can be trained exclusively on pairs of noisy images without any access to ground truth data. For this reason, it is especially useful in biological imaging where it is often the case that trade-offs dictate that ground truth data can't ever be obtained.

However, paired noisy images aren't always easy to obtain, so there was interest in developing methods that could denoise on unpaired training sets of noisy images from some desired domain. The first method capable of this without having sensitive hyperparameters was Noise2Void \cite{Krull_2019_CVPR}. Noise2Void works by training the network to learn a mapping from the noisy image back to itself, masking the center of each receptive field so as to avoid learning the identity.

This basic model of masking the input is known as a blind-spot network, and was heavily refined and expanded upon in \cite{Laine} and much more recently applied in BP-AIDE \cite{BPAIDE} in a manner that is specifically tailored to gaussian-poisson noise. In \cite{FBI} they demonstrate a retooled version of BP-AIDE with much faster inference time.  

A recently developed alternative to blind-spot networks is Neighbor2Neighbor \cite{Ne2Ne} which underlies the method we present in this paper. Neighbor2Neighbor learns to map adjacent pixels in the image to one-another, with the idea being that, except in the most highly dynamic regions of the image, adjacent pixels tend to have a similar underlying signal.

Ultimately, all methods listed in this section require a representative training set of noisy images to train on before being applied. Although we can fairly easily extend some of them to apply to single noisy images without any additional outside information (which we do in our comparisons), in the next section we describe methods that were specifically developed with this task in mind. 

\textbf{Single-image methods:} 
The first method that directly applied itself to the task of single image denoising is Noise2Self \cite{pmlr-v97-batson19a}. Noise2Self is a very similar method to Noise2Void that achieves slighlty better performance, and includes a very thorough mathematical justification for the principles underlying the success of masking based denoising techniques.

Self2Self \cite{Quan_2020_CVPR} was the first single-image method whose performance approaches fully trained methods. Self2Self is a blind-spot method, however instead of replacing masked pixels, it ignores them altogether by using partial convolutions \cite{PConv}. Self2Self also introduces the innovative step of adding dropout and averaging across multiple runs of the same image. However, this comes at a high computational cost, at least under modern hardware constraints.

A very recently published single-image denoiser is R2R \cite{recor}, which achieves even better single-image denoising results than Self2Self on real world images. R2R is quite different than the blindspot network approaches in that it attempts to corrupt single noisy images into noisy image pairs, and then apply a Noise2Noise-like network.

All of the above methods make few assumptions and run out-of-the-box on most single noisy images (with the possible excpetion of R2R \cite{recor}, which we have not tested as it is quite recent and there is no publicly available code yet). In the next section, we present single image methods that are not quite as generically applicable.

\textbf{Single image methods with sensitive hyperparameters:} NL-means \cite{NLM} is one of the easiest and most intuitive non-learning based ways to denoise single images. It denoises by taking the weighted average of all pixels in an image based on how similar we would expect that pixel to be to the target (determined by comparing small square patches centered at those pixels). NL-means is however highly sensitive to a filtering parameter that must be specified by the user for optimal performance.

A similar method is BM3D \cite{4271520}. Since its introduction, BM3D has been one of the gold standards for pure Gaussian noise. It works by unfolding the image into interleaved square patches, clustering those patches based on similarity, and then filtering them before reconstructing the image. BM3D however is not blind and takes, as a parameter, an estimate of the standard deviation of the underlying noise. Moreover, BM3D does not perform well (and was not designed to perform well) on poisson noise.

A much more recent learning based method is Deep Image Prior (DIP) \cite{Ulyanov_2018_CVPR}. DIP works by taking a neural network with randomly initialized weights, and training to reconstruct the noisy image. Similar to Noise2Noise, it will fail to learn the underlying noise (at least at first) and instead learn to denoise the signal. DIP is highly sensitive to the number of iterations, and will quickly overfit if trained too long, for this reason it isn't completely practical as a blind denoiser. For our experiments, we force it to be blind by using a fixed iteration number, however the results it attains are far below what a non-blind version of this algorithm can reach.

\begin{figure*}
\begin{center}
\includegraphics[width=1\linewidth]{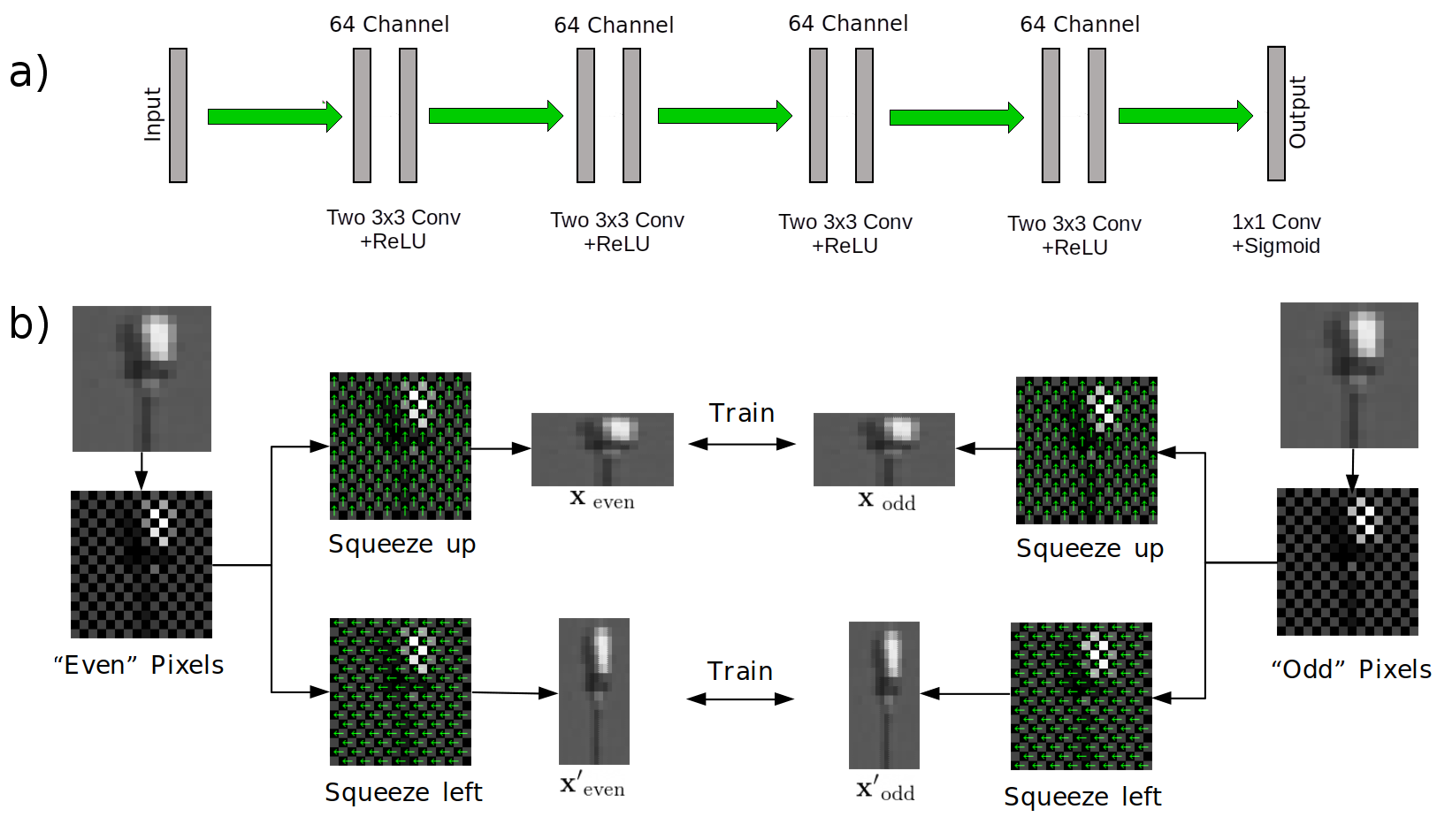}
\end{center}
   \caption{\textbf{a)} Our simple feed-forward CNN architecture. Inputs can be multi-channel, however for best results outputs are always single channel (for rgb images we predict each channel separately). \textbf{b)} Overview of our training scheme. Our neural network learns mappings between pairs of checkerboard downsampled images, each generated from different group of pixels.}
\end{figure*}

\subsubsection*{Results}
\begin{figure*}
\begin{center}
\includegraphics[width=1\linewidth]{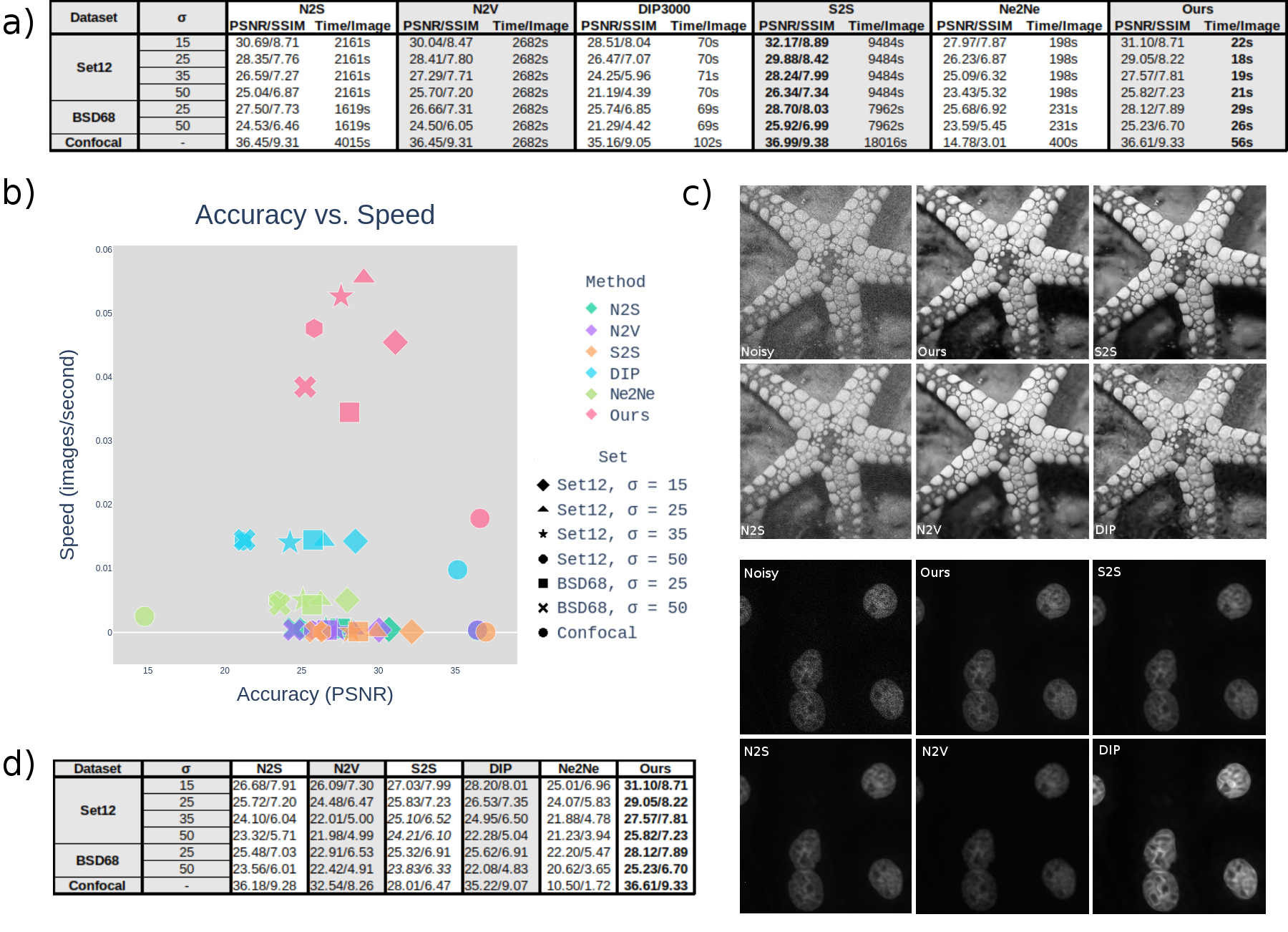}
\end{center}
   \caption{\textbf{a)} Accuracy and per-image time required to denoise on an RTX 5000 mobile GPU, for each dataset using each method. \textbf{b)} Graph of speed (in images per second) versus accuracy (PSNR) of each method on each dataset. \textbf{c)} Visual comparison of each method on starfish image from Set12 and on BPAE cells from Confocal dataset. \textbf{d)} Performance reached by each method, by the time Noise2Fast has completed its denoising. }
\end{figure*}

We demonstrate the speed and accuracy of our method on both simulated Gaussian noise and on real world microscopy data that is subject to Gaussian-Poisson noise. We also benchmark our method on the BSD68 dataset \cite{BSD68} with synthetic Gaussian noise added. See Methods for details on datasets and compared algorithms.

The benchmarking of reference datasets was carried out using a single laptop GPU (RTX 5000 mobile GPU) to better approximate the modest (although still powerful) computational capabilities of the average end user. However, because of the massive amount of time required to test Self2Self on 68 images under these constraints, we rely on their previously published accuracy for comparison and estimate time per image using a random sample of 5 images for this dataset only. On all other datasets (Set12 and Confocal), we run Self2Self on the entire set to obtain accuracy.

On synthetic Gaussian noise our method outperforms everything except Self2Self, which beats us by an average of about 0.7 PSNR across Set12 and BSD68 (see Figure 3a). We did not test our method on the very recent R2R \cite{recor} because there is no publicly available code for this yet, however we note that in the paper the authors assert that their method takes about 30 minutes to run on a $512 \times 512 \times 3$ image using unstated hardware. Therefore, we feel comfortable saying that Noise2Fast is faster than all competing methods, by a significant margin in every case except for DIP3000 (which is far less accurate than ours). 

We also tested our method on Confocal microscopy images, where again our method is considerably faster (300 times) although slightly less accurate than Self2Self. Visual comparison of the results (see Figure 3c) indicate that Noise2Fast appears to smooth the image less than the other methods, creating a more textured look. All methods performed very similarly for the Confocal microscopy dataset, except for DIP3000, which likely needed more iterations to converge, and Neighbor2Neighbor, which seems to not really be suited to single image denoising (nor was it ever intended to be, we adapted it for single image denoising and included it in our benchmarks simply because it is the method most
comparable to our own).

Because 'speed' is just a reflection of the maximum number of iterations we allow each method to run (a parameter we borrow from their published code where possible), we also compared the accuracy of each method if we set the maximum number of iterations so that each program only runs for as long as Noise2Fast takes to fully denoise the image (see Figure 3d). In this case, it is easy to see that no competitor even approaches the accuracy Noise2Fast can achieve in such a short amount of time.

Next we determined the performance of Noise2Fast on larger image datasets of both fixed and live cells acquired on our imaging systems. For this, MDA-MB 231 cells were fixed and either stained for Actin and DNA or endogenously tagged with H33B-mScarlet and mNeon-ACTB. The performance was compared using two different imaging modalities: epifluorescence for the fixed cells and resonance scanning confocal microscopy for the live cells. Based on the linearity of the intensity measurements of our imaging system, our results indicate that we can achieve relatively clear images while exposing our images to ~400 times less light (see Figure 4). Although S2S achieves similar, if not slightly better results, processing time was significantly longer, specifically S2S required 596 core days vs 0.7 for our method to process the video in Fig 4b) on a Tesla V100.

\begin{figure*}
\begin{center}
\includegraphics[width=1\linewidth]{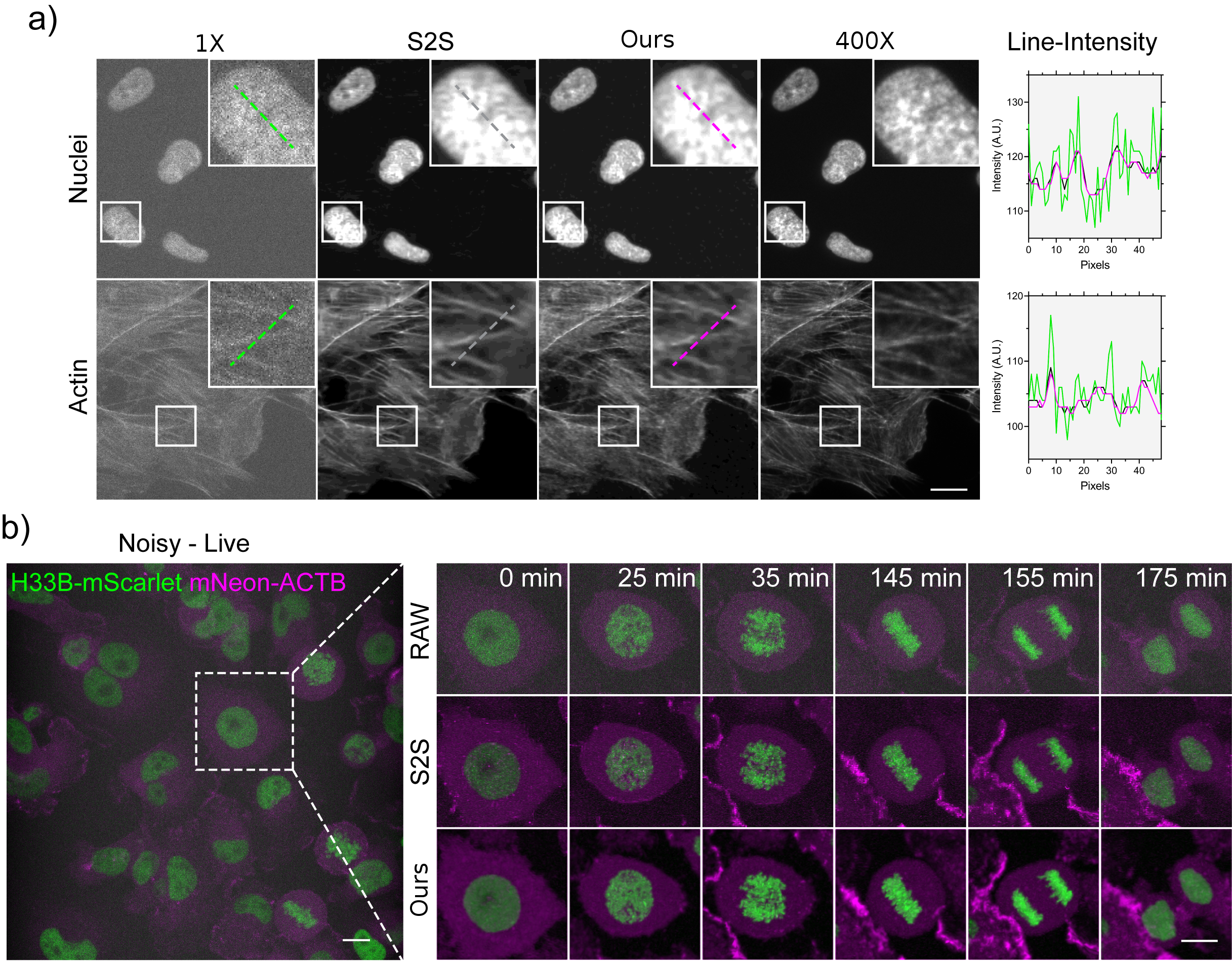}
\end{center}
    \caption{Performance of Noise2Fast on our own microscopy images. \textbf{a)} Comparison of Noise2Fast and Self2Self on epifluorescence images of actin and nuclei in RPE-1 cells with corresponding line intensity profiles. \textbf{b)} Comparison of live confocal imaging of endogenously tagged nuclei (H3-3B-mScarlet) and actin (mNeon-ACTB) in MDA-231 cells. Bars = $10$ $\mu m$.}
\end{figure*}

\subsubsection*{Conclusion}
We proposed Noise2Fast a blind single-image denoiser that rapidly converges to accurate results using only the input image to train on. Our key innovation is building a small discrete training set based on checkerboard downsampling that enables our network to quickly converge. We can monitor the progress training using original noisy image as validation. The accuracy of our method surpasses all but one tested blind single image denoising method, namely Self2Self, however Self2Self takes well over 100 times longer to run and is therefore impractical in situations where fast results are desired, such as in high-throughput and automated-microscopy based pipelines. 
To this end, we hope our very fast and accurate method will integrate well with other AI-based smart microscopy pipelines, such as genome-wide single-cell phenotypic screens \cite{Yan2021}, where denoising will be the initial step in improving the ease and speed of downstream analysis. We also hope that the speed and generality of our method will make it attractive to anyone as a quick, near real-time denoising solution applicable to large live-cell volumetric datasets. 
Additionally, we believe the observed superiority of checkerboard downsampling over traditional $2 \times 2$ downsampling is noteworthy, and the implications this has for full dataset based denoising methods such as Neighbor2Neighbor might be a worthwhile subject of future research.

\section*{Methods}
\subsubsection*{Noise2Fast Implementation Details}

Here, we outline the specifics of our neural network and training scheme, giving the implementation details of the process outlined earlier.

For our neural network, we use a simple feed forward architecture which we explain briefly here, and illustrate in Figure 2a. We start by performing two 32
channel 3x3 convolutions with ReLU activation. We repeat this step three more times, each time doubling the number of channels. In the final step, we do 1x1 convolution followed by sigmoid activation. 

In our initial testing we found that this much simpler architecture outperformed the classical U-net architecture used in the original Noise2Noise paper \cite{pmlr-v80-lehtinen18a}. Although the results aren't that sensitive to the number of hidden layers, we do find a noticeable, albeit small, drop in performance as we add more to our current model. A possible reason for this is that it causes our network to overfit the data much too quickly. This architecture is similar in its simplicity to DnCNN, one major difference being our lack of batch normalization

The main novelty of our method is how we train it. Consider a 2D image $\bf{x} \in \mathbb{R}^{mxn}$. Recall from the theoretical background that we can divide our image in two by using checkerboard downsampling. By taking the even or odd pixels  and squeezing them up to fill in the spaces, as depicted in Figure 1, we can generate two downsampled $m \times \frac{1}{2}n$ images
\begin{align} \mathbf{x}_\text{even}(i,j) &= \mathbf{x}(i,2j+(i \;\mathrm{mod}\; 2)), \\
\mathbf{x}_\text{odd}(i,j) &= \mathbf{x}(i,2j+(i \;\mathrm{mod}\; 2)+1). \end{align}
We can call these the "up" checkerboard downsamples. Notice that we can also squeeze the pixels left to generate two $\frac{1}{2}m \times n$ images
\begin{align} \mathbf{x}_\text{even}'(i,j) &= \mathbf{x}(2i+(j \;\mathrm{mod}\; 2),j), \\ 
\mathbf{x}_\text{odd}'(i,j) &= \mathbf{x}(2i+(j \;\mathrm{mod}\; 2)+1,j). \end{align}
Giving us the "left" checkerboard downsamples. Using these we construct the following four image-pair training set (see Figure 2b for an overview of our training scheme):

\begin{table}[!htbp]
\begin{center}
\begin{tabular}{l|c}
Input & Target \\
\hline
$\mathbf{x}_\text{even}$ & $\mathbf{x}_\text{odd}$ \\
$\mathbf{x}_\text{odd}$ & $\mathbf{x}_\text{even}$ \\
$\mathbf{x}_\text{even}'$ & $\mathbf{x}_\text{odd}'$ \\
$\mathbf{x}_\text{odd}'$ & $\mathbf{x}_\text{even}'$ \\
\end{tabular}
\end{center}
\end{table}
We feed this training data one-by-one into our neural network (batch size = 1). At each iteration we compute the binary cross-entropy (BCE) loss between the target and the output of our neural network, and adjust our weights using the Adam optimizer \cite{kingma2014method} with learning rate set to $0.001$. For validation, after each epoch we run the original full sized noisy image through our network and compute the mean squared error (MSE) between the noisy input image and the "denoised" output of our network. 

We validate this way because our initial testing on images with known ground truths showed that the MSE between the denoised image and the ground truth image plateaued at roughly the same time as the MSE between the denoised image and the original noisy image. After this point, results start to get worse, much like the way DIP starts to overfit at a certain point, however in our case we can use this validation protocol to prevent that without introducing a sensitive case-dependant parameter. If one hundred epochs have passed without any improvement to the best validation score, we terminate the program, and output the average of the last one hundred validation tests as the denoised image.
One unusual feature of our training scheme is our usage of a BCE loss function usually reserved for classification based tasks. Our main motivation for using this loss function is to deal with class imbalance. When an image contains only a small object on a black background, it will sometimes just learn to map everything to black. We found that using BCE loss fixed this in all cases we could find, without affecting overall performance as compared to MSE loss. We also tried binary focal loss \cite{lin2018focal}, but the results were not as good.

\subsubsection*{Compared Datasets}
For blind Gaussian denoising we use use the grayscale BSD68 \cite{BSD68} dataset, as was used in \cite{Krull_2019_CVPR} and a multitude of other denoising papers. BSD68 consists of 68 clear 481x321 photographs to which we add synthetic gaussian noise. However, to show the effect of spatial resolution on speed and performance, we additionally tested the methods on Set12 which contains a mixture of 256x256 and 512x512 images.

For performance on real world confocal microscopy, we used a subset of the confocal microscopy images in Fluorescent Microscopy Dataset (FMD) \cite{Confoc5} that we refer to as "Confocal". This dataset contains, among other things, images of biological materials such as cells, zebrafish, and mouse brain tissues acquired using commercial confocal microscopes. As described in their paper, ground truth values are estimated by averaging together all 60000 noisy images in a given set.

\subsubsection*{Compared Methods}
We compare denoising and speed performance against five other blind single image denoisers: Noise2Self \cite{pmlr-v97-batson19a} (N2S), Noise2Void \cite{Krull_2019_CVPR}(N2V), Self2Self \cite{Quan_2020_CVPR}(S2S), Neighbor2Neighbor \cite{Ne2Ne}(Ne2Ne) and Deep-Image Prior \cite{Ulyanov_2018_CVPR}(DIP). Not all of these methods were originally designed for single image denoising. We will describe how we configured each of these methods in turn, we adhere to published code as much as possible.

\textbf{Self2Self:} For S2S we use the default published settings of 150000 iterations and a learning rate of 1e-4. We standardize our images differently than S2S and some of these other methods. For example, we do not clip our input noisy data [0,255] at any point. To account for this difference, we have rewritten the dataloaders for Self2Self and other methods to ensure consistency of comparison.

\textbf{Noise2Self:} For N2S the only change we make from their published single shot denoising notebook is to increase the number of iterations from 500 to 20000, as we found that 500 iterations wasn't nearly enough to achieve good results on these datasets.

\textbf{Noise2Void:} For N2V we found that their ImageJ plugin worked much better than their GitHub code for single image denoising. We therefore used the imageJ version for benchmarking purposes, which is why our results on this method deviate so much from previous publications. We used a patch size of 64 with 100 epochs and 100 steps per epoch, a batch size of 16 per step, and a neighborhood radius of 5.

\textbf{DIP:} If we fix the maximum number of iterations, DIP becomes a blind denoiser. However, as noted in \cite{Quan_2020_CVPR}, it performs better as a non-blind denoiser. For comparison purposes however, we will set the maximum number of iterations at 3000, as the authors of DIP have done in their example code on GitHub, and call this DIP3000. This turns it into a blind single shot denoiser, fully comparable in scope to our method.

\textbf{Neighbor2Neighbor:} For Neighbor2Neighbor we used the adaptation of the code found here: https://github.com/neeraj3029/Ne2Ne-Image-Denoising. We adapted the script to single image denoising and attempted in good-faith to optimize for the task as best we could, however we found that the results were inconsistent. We believe that this method is probably best suited to datasets as the authors intended, and not single images. We include these results only to illustrate the need to change Neighbor2Neighbor in order to achieve fast and accurate single image denoising results, as we have done in this paper. We do not believe our results are a fair illustration of the power of Neighbor2Neighbor when applied to the tasks it was designed for and we have therefore excluded it from our visual illustrations. We used a learning rate of 0.0003 and trained for 100 epochs, as suggested in their paper for synthetic datasets.

\subsubsection*{Fluorescence Microscopy Images}
For fixed immunofluorescence microscopy, RPE-1 cells were fixed with 4\% paraformaldehyde at room temperature for 10 min. The cells were then blocked with a blocking buffer (5\% BSA and 0.5\% Triton X-100 in PBS) for 30 min. Cells were washed with PBS and subsequently incubated with phalloidin-Alexa488 (Molecular Probes) and DAPI in blocking solution for 1 hour. After a final wash with PBS, the coverslips were mounted on glass slides by inverting them onto mounting solution (ProLong Gold antifade; Molecular Probes). For the fixed imaging in Figure 4A, single Z slices of cells were imaged using Nikon Ti2E/CREST X-Light V2 LFOV25 spinning disk confocal microscope in widefield mode using a 60×/1.4 NA oil-immersion Plan-Apochromat lambda objective. The microscope was outfitted with a Photometrics Prime95B 25mm FOV ultra-high sensitivity sCMOS camera and images were captured with no binning using the full 25mm diagonal FOV area at 1608px by 1608px with a bit depth of 16bit. After capture, 500px by 500px areas were cropped and used as our input dataset. For live imaging in Figure 4B, endogenously tagged MDA-MB 231 cells were seeded in Nunc Lab-Tek Chamber Slides and imaged on the Nikon Ti2E/AIR-HD25 scanning confocal microscope with temperature and CO2 control, using a 40×/1.15 NA water-immersion objective Apochromat lambda S objective. High-speed image acquisition was carried out with the resonance scan head with 2x averaging at 1024px by 1024px. Full volumes of cells were captured (Z total = $20$ $\mu m$, Z interval = $0.5$ $\mu m$) every 5 minutes for 24 hours. Images were denoised as individual Z-slices and max projected. All are displayed with auto scaled LUTs.

\subsubsection*{Ablation Study}
For our ablation study, we compare three different refinements of the model. First, we replace our unusual checkerboard downsampling with a more conventional downsample where we divide our image into $2 \times 2$ blocks used in Neighbor2Neighbor, and create four images consisting of all the top-left, top-right, bottom-left and bottom-right pixels, respectively. This has the advantage of preserving the proportions of our original image as well the structure of distances between pixels, however despite this advantage it does not perform as well (Table 1 - Quad). Second, we replace our feed forward neural network with a U-net architecture, which is the standard network used in Self2Self and Noise2Void. Again, our performance drops (Table 1 - Unet). Finally, using known ground truth values, we manually subtract out the $\mathbf{s}_\text{odd} - \mathbf{s}_\text{even}$ term in $\ref{eqn:ODD}$ and show that this has virtually no impact on our denoising results, hence this term is not having a significant impact on our algorithm (Table 1 - Exact).

\begin{table}[!htbp]
\begin{center}

\begin{tabular}{|l|c c c c|}
\hline
Dataset & Normal & Quad & Unet & Exact \\
\hline
BSD68, $\sigma = 25$ & 28.12 & 27.56 & 27.76 & 28.12 \\
BSD68, $\sigma = 50$ & 25.23 & 24.97 & 25.01 & 25.24 \\
\hline
\end{tabular}
\end{center}
\caption{PSNR on BSD68 dataset for variants of Noise2Fast.}
\end{table}

\subsubsection*{Code Availabilty}
Our code is publicly available at the following URL: https://github.com/pelletierlab/Noise2Fast

{\small
\bibliographystyle{naturemag}

\begin{thebibliography}{10}
\expandafter\ifx\csname url\endcsname\relax
  \def\url#1{\texttt{#1}}\fi
\expandafter\ifx\csname urlprefix\endcsname\relax\def\urlprefix{URL }\fi
\providecommand{\bibinfo}[2]{#2}
\providecommand{\eprint}[2][]{\url{#2}}

\bibitem{Confoc5}
\bibinfo{author}{Zhang, Y.} \emph{et~al.}
\newblock \bibinfo{title}{A poisson-gaussian denoising dataset with real
  fluorescence microscopy images}.
\newblock In \emph{\bibinfo{booktitle}{CVPR}} (\bibinfo{year}{2019}).

\bibitem{Zhang_2017}
\bibinfo{author}{Zhang, K.}, \bibinfo{author}{Zuo, W.}, \bibinfo{author}{Chen,
  Y.}, \bibinfo{author}{Meng, D.} \& \bibinfo{author}{Zhang, L.}
\newblock \bibinfo{title}{Beyond a gaussian denoiser: Residual learning of deep
  cnn for image denoising}.
\newblock \emph{\bibinfo{journal}{IEEE Transactions on Image Processing}}
  \textbf{\bibinfo{volume}{26}}, \bibinfo{pages}{3142–3155}
  (\bibinfo{year}{2017}).
\newblock \urlprefix\url{http://dx.doi.org/10.1109/TIP.2017.2662206}.

\bibitem{pmlr-v80-lehtinen18a}
\bibinfo{author}{Lehtinen, J.} \emph{et~al.}
\newblock \bibinfo{title}{{N}oise2{N}oise: Learning image restoration without
  clean data}.
\newblock In \bibinfo{editor}{Dy, J.} \& \bibinfo{editor}{Krause, A.} (eds.)
  \emph{\bibinfo{booktitle}{Proceedings of the 35th International Conference on
  Machine Learning}}, vol.~\bibinfo{volume}{80} of
  \emph{\bibinfo{series}{Proceedings of Machine Learning Research}},
  \bibinfo{pages}{2965--2974} (\bibinfo{publisher}{PMLR},
  \bibinfo{address}{Stockholmsmässan, Stockholm Sweden},
  \bibinfo{year}{2018}).
\newblock \urlprefix\url{http://proceedings.mlr.press/v80/lehtinen18a.html}.

\bibitem{Krull_2019_CVPR}
\bibinfo{author}{Krull, A.}, \bibinfo{author}{Buchholz, T.-O.} \&
  \bibinfo{author}{Jug, F.}
\newblock \bibinfo{title}{Noise2void - learning denoising from single noisy
  images}.
\newblock In \emph{\bibinfo{booktitle}{Proceedings of the IEEE/CVF Conference
  on Computer Vision and Pattern Recognition (CVPR)}} (\bibinfo{year}{2019}).

\bibitem{pmlr-v97-batson19a}
\bibinfo{author}{Batson, J.} \& \bibinfo{author}{Royer, L.}
\newblock \bibinfo{title}{{N}oise2{S}elf: Blind denoising by self-supervision}.
\newblock In \bibinfo{editor}{Chaudhuri, K.} \& \bibinfo{editor}{Salakhutdinov,
  R.} (eds.) \emph{\bibinfo{booktitle}{Proceedings of the 36th International
  Conference on Machine Learning}}, vol.~\bibinfo{volume}{97} of
  \emph{\bibinfo{series}{Proceedings of Machine Learning Research}},
  \bibinfo{pages}{524--533} (\bibinfo{publisher}{PMLR}, \bibinfo{year}{2019}).
\newblock \urlprefix\url{http://proceedings.mlr.press/v97/batson19a.html}.

\bibitem{Quan_2020_CVPR}
\bibinfo{author}{Quan, Y.}, \bibinfo{author}{Chen, M.}, \bibinfo{author}{Pang,
  T.} \& \bibinfo{author}{Ji, H.}
\newblock \bibinfo{title}{Self2self with dropout: Learning self-supervised
  denoising from single image}.
\newblock In \emph{\bibinfo{booktitle}{Proceedings of the IEEE/CVF Conference
  on Computer Vision and Pattern Recognition (CVPR)}} (\bibinfo{year}{2020}).

\bibitem{recor}
\bibinfo{author}{Pang, T.}, \bibinfo{author}{Zheng, H.}, \bibinfo{author}{Quan,
  Y.} \& \bibinfo{author}{Ji, H.}
\newblock \bibinfo{title}{Recorrupted-to-recorrupted: Unsupervised deep
  learning for image denoising}.
\newblock In \emph{\bibinfo{booktitle}{Proceedings of the IEEE/CVF Conference
  on Computer Vision and Pattern Recognition (CVPR)}},
  \bibinfo{pages}{2043--2052} (\bibinfo{year}{2021}).

\bibitem{Ne2Ne}
\bibinfo{author}{Huang, T.}, \bibinfo{author}{Li, S.}, \bibinfo{author}{Jia,
  X.}, \bibinfo{author}{Lu, H.} \& \bibinfo{author}{Liu, J.}
\newblock \bibinfo{title}{Neighbor2neighbor: Self-supervised denoising from
  single noisy images}.
\newblock In \emph{\bibinfo{booktitle}{Proceedings of the IEEE/CVF Conference
  on Computer Vision and Pattern Recognition (CVPR)}},
  \bibinfo{pages}{14781--14790} (\bibinfo{year}{2021}).

\bibitem{5995401}
\bibinfo{author}{{Zontak}, M.} \& \bibinfo{author}{{Irani}, M.}
\newblock \bibinfo{title}{Internal statistics of a single natural image}.
\newblock In \emph{\bibinfo{booktitle}{CVPR 2011}}, \bibinfo{pages}{977--984}
  (\bibinfo{year}{2011}).

\bibitem{5459271}
\bibinfo{author}{{Glasner}, D.}, \bibinfo{author}{{Bagon}, S.} \&
  \bibinfo{author}{{Irani}, M.}
\newblock \bibinfo{title}{Super-resolution from a single image}.
\newblock In \emph{\bibinfo{booktitle}{2009 IEEE 12th International Conference
  on Computer Vision}}, \bibinfo{pages}{349--356} (\bibinfo{year}{2009}).

\bibitem{7299621}
\bibinfo{author}{{Zhang}, Y.}, \bibinfo{author}{{Ling}, F.},
  \bibinfo{author}{{Li}, X.} \& \bibinfo{author}{{Du}, Y.}
\newblock \bibinfo{title}{Super-resolution land cover mapping using multiscale
  self-similarity redundancy}.
\newblock \emph{\bibinfo{journal}{IEEE Journal of Selected Topics in Applied
  Earth Observations and Remote Sensing}} \textbf{\bibinfo{volume}{8}},
  \bibinfo{pages}{5130--5145} (\bibinfo{year}{2015}).

\bibitem{Hasle2020}
\bibinfo{author}{Hasle, N.} \emph{et~al.}
\newblock \bibinfo{title}{High-throughput, microscope-based sorting to dissect
  cellular heterogeneity}.
\newblock \emph{\bibinfo{journal}{Molecular Systems Biology}}
  \textbf{\bibinfo{volume}{16}} (\bibinfo{year}{2020}).
\newblock \urlprefix\url{https://doi.org/10.15252/msb.20209442}.

\bibitem{Kanfer2021}
\bibinfo{author}{Kanfer, G.} \emph{et~al.}
\newblock \bibinfo{title}{Image-based pooled whole-genome {CRISPRi} screening
  for subcellular phenotypes}.
\newblock \emph{\bibinfo{journal}{Journal of Cell Biology}}
  \textbf{\bibinfo{volume}{220}} (\bibinfo{year}{2021}).
\newblock \urlprefix\url{https://doi.org/10.1083/jcb.202006180}.

\bibitem{Yan2021}
\bibinfo{author}{Yan, X.} \emph{et~al.}
\newblock \bibinfo{title}{High-content imaging-based pooled {CRISPR} screens in
  mammalian cells}.
\newblock \emph{\bibinfo{journal}{Journal of Cell Biology}}
  \textbf{\bibinfo{volume}{220}} (\bibinfo{year}{2021}).
\newblock \urlprefix\url{https://doi.org/10.1083/jcb.202008158}.

\bibitem{firstDn}
\bibinfo{author}{Jain, V.} \& \bibinfo{author}{Seung, S.}
\newblock \bibinfo{title}{Natural image denoising with convolutional networks}.
\newblock In \bibinfo{editor}{Koller, D.}, \bibinfo{editor}{Schuurmans, D.},
  \bibinfo{editor}{Bengio, Y.} \& \bibinfo{editor}{Bottou, L.} (eds.)
  \emph{\bibinfo{booktitle}{Advances in Neural Information Processing
  Systems}}, vol.~\bibinfo{volume}{21}, \bibinfo{pages}{769--776}
  (\bibinfo{publisher}{Curran Associates, Inc.}, \bibinfo{year}{2009}).
\newblock
  \urlprefix\url{https://proceedings.neurips.cc/paper/2008/file/c16a5320fa475530d9583c34fd356ef5-Paper.pdf}.

\bibitem{OtherDn}
\bibinfo{author}{Mao, X.}, \bibinfo{author}{Shen, C.} \& \bibinfo{author}{Yang,
  Y.-B.}
\newblock \bibinfo{title}{Image restoration using very deep convolutional
  encoder-decoder networks with symmetric skip connections}.
\newblock In \bibinfo{editor}{Lee, D.}, \bibinfo{editor}{Sugiyama, M.},
  \bibinfo{editor}{Luxburg, U.}, \bibinfo{editor}{Guyon, I.} \&
  \bibinfo{editor}{Garnett, R.} (eds.) \emph{\bibinfo{booktitle}{Advances in
  Neural Information Processing Systems}}, vol.~\bibinfo{volume}{29},
  \bibinfo{pages}{2802--2810} (\bibinfo{publisher}{Curran Associates, Inc.},
  \bibinfo{year}{2016}).
\newblock
  \urlprefix\url{https://proceedings.neurips.cc/paper/2016/file/0ed9422357395a0d4879191c66f4faa2-Paper.pdf}.

\bibitem{FFD}
\bibinfo{author}{{Zhang}, K.}, \bibinfo{author}{{Zuo}, W.} \&
  \bibinfo{author}{{Zhang}, L.}
\newblock \bibinfo{title}{Ffdnet: Toward a fast and flexible solution for
  cnn-based image denoising}.
\newblock \emph{\bibinfo{journal}{IEEE Transactions on Image Processing}}
  \textbf{\bibinfo{volume}{27}}, \bibinfo{pages}{4608--4622}
  (\bibinfo{year}{2018}).

\bibitem{Sub-pixel}
\bibinfo{author}{Shi, W.} \emph{et~al.}
\newblock \bibinfo{title}{Real-time single image and video super-resolution
  using an efficient sub-pixel convolutional neural network}.
\newblock In \emph{\bibinfo{booktitle}{2016 {IEEE} Conference on Computer
  Vision and Pattern Recognition ({CVPR})}} (\bibinfo{publisher}{{IEEE}},
  \bibinfo{year}{2016}).
\newblock \urlprefix\url{https://doi.org/10.1109/cvpr.2016.207}.

\bibitem{Laine}
\bibinfo{author}{Laine, S.}, \bibinfo{author}{Karras, T.},
  \bibinfo{author}{Lehtinen, J.} \& \bibinfo{author}{Aila, T.}
\newblock \bibinfo{title}{High-quality self-supervised deep image denoising}.
\newblock In \bibinfo{editor}{Wallach, H.} \emph{et~al.} (eds.)
  \emph{\bibinfo{booktitle}{Advances in Neural Information Processing
  Systems}}, vol.~\bibinfo{volume}{32} (\bibinfo{publisher}{Curran Associates,
  Inc.}, \bibinfo{year}{2019}).
\newblock
  \urlprefix\url{https://proceedings.neurips.cc/paper/2019/file/2119b8d43eafcf353e07d7cb5554170b-Paper.pdf}.

\bibitem{BPAIDE}
\bibinfo{author}{Byun, J.} \& \bibinfo{author}{Moon, T.}
\newblock \bibinfo{title}{Learning blind pixelwise affine image denoiser with
  single noisy images}.
\newblock \emph{\bibinfo{journal}{IEEE Signal Processing Letters}}
  \textbf{\bibinfo{volume}{27}}, \bibinfo{pages}{1105--1109}
  (\bibinfo{year}{2020}).

\bibitem{FBI}
\bibinfo{author}{Byun, J.}, \bibinfo{author}{Cha, S.} \& \bibinfo{author}{Moon,
  T.}
\newblock \bibinfo{title}{Fbi-denoiser: Fast blind image denoiser for
  poisson-gaussian noise}.
\newblock In \emph{\bibinfo{booktitle}{Proceedings of the IEEE/CVF Conference
  on Computer Vision and Pattern Recognition (CVPR)}},
  \bibinfo{pages}{5768--5777} (\bibinfo{year}{2021}).

\bibitem{PConv}
\bibinfo{author}{Liu, G.} \emph{et~al.}
\newblock \bibinfo{title}{Image inpainting for irregular holes using partial
  convolutions}.
\newblock \emph{\bibinfo{journal}{CoRR}}
  \textbf{\bibinfo{volume}{abs/1804.07723}} (\bibinfo{year}{2018}).
\newblock \urlprefix\url{http://arxiv.org/abs/1804.07723}.
\newblock \eprint{1804.07723}.

\bibitem{NLM}
\bibinfo{author}{Buades, A.}, \bibinfo{author}{Coll, B.} \&
  \bibinfo{author}{Morel, J.-M.}
\newblock \bibinfo{title}{A non-local algorithm for image denoising}.
\newblock In \emph{\bibinfo{booktitle}{2005 {IEEE} Computer Society Conference
  on Computer Vision and Pattern Recognition ({CVPR}{\textquotesingle}05)}}
  (\bibinfo{publisher}{{IEEE}}).
\newblock \urlprefix\url{https://doi.org/10.1109/cvpr.2005.38}.
\bibitem{4271520}
\bibinfo{author}{{Dabov}, K.}, \bibinfo{author}{{Foi}, A.},
  \bibinfo{author}{{Katkovnik}, V.} \& \bibinfo{author}{{Egiazarian}, K.}
\newblock \bibinfo{title}{Image denoising by sparse 3-d transform-domain
  collaborative filtering}.
\newblock \emph{\bibinfo{journal}{IEEE Transactions on Image Processing}}
  \textbf{\bibinfo{volume}{16}}, \bibinfo{pages}{2080--2095}
  (\bibinfo{year}{2007}).

\bibitem{Ulyanov_2018_CVPR}
\bibinfo{author}{Ulyanov, D.}, \bibinfo{author}{Vedaldi, A.} \&
  \bibinfo{author}{Lempitsky, V.}
\newblock \bibinfo{title}{Deep image prior}.
\newblock In \emph{\bibinfo{booktitle}{Proceedings of the IEEE Conference on
  Computer Vision and Pattern Recognition (CVPR)}} (\bibinfo{year}{2018}).

\bibitem{BSD68}
\bibinfo{author}{Martin, D.}, \bibinfo{author}{Fowlkes, C.},
  \bibinfo{author}{Tal, D.} \& \bibinfo{author}{Malik, J.}
\newblock \bibinfo{title}{A database of human segmented natural images and its
  application to evaluating segmentation algorithms and measuring ecological
  statistics}.
\newblock In \emph{\bibinfo{booktitle}{Proc. 8th Int'l Conf. Computer Vision}},
  vol.~\bibinfo{volume}{2}, \bibinfo{pages}{416--423} (\bibinfo{year}{2001}).

\bibitem{kingma2014method}
\bibinfo{author}{Kingma, D.~P.} \& \bibinfo{author}{Ba, J.}
\newblock \bibinfo{title}{Adam: A method for stochastic optimization}
  (\bibinfo{year}{2014}).
\newblock \urlprefix\url{http://arxiv.org/abs/1412.6980}.
\newblock.

\bibitem{lin2018focal}
\bibinfo{author}{Lin, T.-Y.}, \bibinfo{author}{Goyal, P.},
  \bibinfo{author}{Girshick, R.}, \bibinfo{author}{He, K.} \&
  \bibinfo{author}{Dollár, P.}
\newblock \bibinfo{title}{Focal loss for dense object detection}
  (\bibinfo{year}{2018}).
\newblock \eprint{1708.02002}.

\end{thebibliography}
}

\end{document}